\documentclass[prb, aps, prl, showpacs, onecolumn]{revtex4}

\usepackage{graphics}
\usepackage{dcolumn}
\usepackage{epsfig}
\begin{document}

\title{\texttt{EPW}: A program for calculating the 
  electron-phonon coupling \\ using maximally localized Wannier functions }

\author{Jesse Noffsinger}
\email{jnoffsinger@civet.berkeley.edu}
\affiliation
{Department of Physics, 
University of California, Berkeley,
California 94720, USA.}
\affiliation
{Materials Sciences Division, 
Lawrence Berkeley National Laboratory, Berkeley, 
California 94720, USA.}

\author{Feliciano Giustino}
\affiliation
{Department of Materials,
University of Oxford,
Parks Road, Oxford, OX1 3PH, UK.}

\author{Brad D. Malone}
\affiliation
{Department of Physics, 
University of California, Berkeley, 
California 94720, USA.}
\affiliation
{Materials Sciences Division, 
Lawrence Berkeley National Laboratory, Berkeley, 
California 94720, USA.}

\author{Cheol-Hwan Park}
\affiliation
{Department of Physics, 
University of California, Berkeley, 
California 94720, USA.}
\affiliation
{Materials Sciences Division, 
Lawrence Berkeley National Laboratory, Berkeley, 
California 94720, USA.}

\author{Steven G. Louie}
\affiliation
{Department of Physics, 
University of California, Berkeley, 
California 94720, USA.}
\affiliation
{Materials Sciences Division, 
Lawrence Berkeley National Laboratory, Berkeley, 
California 94720, USA.}

\author{Marvin L. Cohen} 
\affiliation
{Department of Physics, 
University of California, Berkeley, 
California 94720, USA.}
\affiliation
{Materials Sciences Division, 
Lawrence Berkeley National Laboratory, Berkeley, 
California 94720, USA.}

\date{\today}

\begin{abstract}
\texttt{EPW} (\underline{E}lectron-\underline{P}honon coupling 
using \underline{W}annier functions) is a program written in 
\texttt{FORTRAN90} for calculating the electron-phonon coupling
in periodic systems using density-functional perturbation theory
and maximally-localized Wannier functions. \texttt{EPW} can 
calculate electron-phonon interaction self-energies, electron-phonon
spectral functions, and total as well as mode-resolved electron-phonon
coupling strengths. The calculation of the electron-phonon coupling 
requires a very accurate sampling of electron-phonon scattering 
processes throughout the Brillouin zone, hence reliable 
calculations can be prohibitively time-consuming. \texttt{EPW} 
combines the Kohn-Sham electronic eigenstates and the vibrational 
eigenmodes provided by the \texttt{Quantum-ESPRESSO} package
\cite{pwscf} with the maximally localized Wannier functions 
provided by the \texttt{wannier90} package \cite{w90} in order to 
generate electron-phonon matrix elements on arbitrarily dense 
Brillouin zone grids using a generalized Fourier interpolation.  
This feature of \texttt{EPW} leads to fast and accurate calculations
of the electron-phonon coupling, and enables the study of
the electron-phonon coupling in large and complex systems.
\end{abstract}


\maketitle

{\bf PROGRAM SUMMARY}

\begin{small}
\noindent
{\em Program Title:} EPW                                      \\
{\em Journal Reference:}                                      \\
{\em Catalogue identifier:}                                   \\
{\em Licensing provisions:} GNU Public License                \\
{\em Programming language:} Fortran 90                        \\
{\em Computer:} any architecture with a Fortran 90 compiler   \\
{\em RAM: } heavily system dependent, as small as a few MB                         \\
{\em Number of processors used:} optimized for 1 to 64 processors \\
{\em Classification:} 7                                         \\
{\em External routines/libraries:} BLAS, LAPACK, MPI, FFTW        \\
{\em Subprograms used:} wannier90                                 \\
{\em Nature of problem:} 
The calculation of the electron-phonon coupling from first principles
requires a very accurate sampling of electron-phonon scattering
processes throughout the Brillouin zone; hence reliable
calculations can be prohibitively time-consuming.\\
{\em Solution method:}
\texttt{EPW} makes use of a real-space formulation and combines the 
Kohn-Sham electronic eigenstates and the vibrational
eigenmodes provided by the \texttt{Quantum-ESPRESSO} package
with the maximally localized Wannier functions
provided by the \texttt{wannier90} package in order to
generate electron-phonon matrix elements on arbitrarily dense
Brillouin zone grids using a generalized Fourier interpolation. \\
{\em Running time:} single processor examples typically take 5-10 minutes \\
{\em References:}
Giustino, F., Cohen, M. L., and Louie, S. G., Physical Review B {\bf 76} (2007) 165108.
\end{small}

\section{Introduction}

The electron-phonon interaction plays a crucial role in the 
electron and lattice dynamics of condensed matter systems. For 
example, phenomena such as the electrical resistivity 
\cite{elec.resis} and conventional superconductivity 
\cite{schrieffer.book} are a direct consequence of the 
interaction between electrons and lattice vibrations.
The electron-phonon interaction also plays an important role 
in the thermoelectric effect \cite{thermoelectric}. Other 
fundamental physical phenomena such as the Kohn effect 
\cite{kohn.effect} and the Peierls \cite{peierls} distortions 
are also direct consequences of the electron-phonon interaction. 
The electron-phonon interaction is also responsible for the 
broadening of the spectral lines in angle-resolved 
photoemission spectroscopy \cite{rmp.arpes} and in vibrational 
spectroscopies \cite{vibra.broadening}, as well as for the 
temperature dependence of the band gaps in semiconductors 
\cite{cardona.bandgap}.

The calculation of the electron-phonon coupling from first-principles
is challenging because of the necessity of evaluating Brillouin zone 
integrals with high accuracy.  Such calculation 
requires the evaluation of matrix elements between electronic 
states connected by phonon wavevectors \cite{pkl.mlc}.  Well-established 
software packages are available for computing electronic states and 
eigenvalues through density-functional theory \cite{dft,dft.method,pwscf}, 
as well as phonon frequencies and eigenmodes through density-functional 
perturbation theory \cite{baroni.dfpt}.  However large numbers of 
matrix elements may be necessary to achieve numerical convergence of 
the Brillouin zone intergrals over these matrix elements. 

For example, in order to compute the electronic lifetimes associated 
with the electron-phonon interaction it is necessary to evaluate a 
Brillouin zone integral over all the possible phonon wavevectors 
(thousands to millions). Since lattice-dynamical calculations for 
each phonon wavevector are at least as expensive as self-consistent 
total energy minimizations, achieving numerical convergence in the 
Brillouin zone integrals over the phonon wavevectors by 
brute-force calculations may become a prohibitive computational
task. \texttt{EPW} exploits the real-space localization of Wannier 
functions to generate large numbers of first-principles 
electron-phonon matrix elements through a generalized Fourier 
interpolation. \texttt{EPW} therefore enables affordable, accurate, 
and extremely efficient calculations of the electron-phonon 
coupling \cite{giustino.method}. The use of maximally localized 
Wannier functions (MLWFs) \cite{marzari.wannier,souza.wannier} to 
calculate Brillouin zone integrals with high accuracy has been the 
object of a number of other studies
\cite{PhysRevB.75.195121,PhysRevB.76.195109,PhysRevB.74.195118,
PhysRevB.79.045109,giustino.diamond,PhysRevLett.99.086804,
PhysRevB.77.113410,PhysRevB.77.180507,PhysRevLett.102.076803,
noffsinger.sic,giustino.nature,chp.nano1,chp.nano2}.

In this communication we outline the functionalities of \texttt{EPW} 
and the details of the technical release (Sec.~II), we review the 
individual sections of the code (Sec.~III), and we describe
its parallel structure (Sec.~VI). We then illustrate the capabilities
of \texttt{EPW} through example calculations (Sec.~V). We finally 
discuss some future directions for additional development (Sec.~VI). 
Some technical details are given in the Appendix (Sec.~VII). 

\section{Functionalities and technical release}

\texttt{EPW} is a program written in \texttt{FORTRAN90} 
which calculates the electron-phonon coupling from first principles
using maximally localized Wannier functions.  \texttt{EPW} uses 
information provided by \texttt{Quantum-ESPRESSO} \cite{pwscf} 
and \texttt{wannier90}~\cite{w90}, and runs as a post-processing tool. 
Electrons are described using density functional theory (DFT) 
\cite{dft,dft.method} with plane-waves  and pseudopotentials, either 
separable norm-conserving \cite{fuchs.pseudo, cohen.scripta, 
tm.pseudo, perdew.zunger} or Vanderbilt ultrasoft \cite{uspp}. Lattice 
dynamical properties are calculated within density functional 
perturbation theory (DFPT) \cite{baroni.dfpt}.  The theoretical 
background and methodology are thoroughly described in Ref. 
\cite{giustino.method}.  Examples of quantities \cite{allenmikovic} which 
can be computed using \texttt{EPW} include:

\begin{itemize}
\item{ the total electron-phonon coupling strength $\lambda$,}
\item{ the phonon self-energy associated with the electron-phonon interaction 
within the Migdal approximation,}
\item{ the electron self-energy associated with the electron-phonon interaction 
within the Migdal approximation,}
\item{ the Eliashberg electron-phonon spectral function  $\alpha^2F$,}
\item{ the transport electron-phonon spectral function $\alpha^2F_{\rm T}$.}
\end{itemize}

\texttt{EPW} is freely available under the GNU General Public 
License (GPL). The current version is developed and maintained 
using Subversion and is accessible to prospective developers and 
end-users at the website \texttt{epw.org.uk}. \texttt{EPW} employs 
the freely available \texttt{FFTW, BLAS, LAPACK} libraries in 
conjunction with several subroutines distributed within the 
\texttt{Quantum-ESPRESSO} package. Several subroutines in 
\texttt{EPW} are based upon modified \texttt{Quantum-ESPRESSO} 
subroutines as permitted under the GPL. The parallelization
is achieved through the \texttt{MPI} library specification for 
message passing. The current version of \texttt{EPW}, v2.3, 
includes approximately 9000 independent lines of \texttt{FORTRAN90} 
code. In addition to the source code, several complete example 
calculations are provided with the \texttt{EPW} distribution. 

The inputs to \texttt{EPW} are as follows:
\begin{itemize}
\item{Phonon dynamical matrices for the wavevectors of a uniform 
  Brillouin-zone grid centered at $\Gamma$ (\texttt{prefix.dyn} 
  files from \texttt{Quantum-ESPRESSO}). Only wavevectors within 
  the irreducible wedge of the Brillouin zone are required.} 
\item{The derivatives of the self-consistent potential with 
  respect to the phonon perturbations, for the same wavevectors 
  as above (\texttt{prefix.dvscf} files from \texttt{Quantum-ESPRESSO}).}
\item{The electron eigenfunctions and eigenvalues for the 
  wavevectors of a uniform Brillouin-zone grid centered at 
  $\Gamma$ (\texttt{prefix.wfc} or \texttt{prefix.save/}   
  files from \texttt{Quantum-ESPRESSO}).} 
\item{Norm-conserving pseudopotentials \cite{cohen.scripta} or 
  Vanderbilt ultrasoft pseudopotentials \cite{uspp},}
\item{A plain text input file specifying the runtime parameters.}
\end{itemize}

The electronic wavefunctions are calculated on a uniform grid using 
\texttt{Quantum-ESPRESSO}. Dynamical matrices and the derivatives 
of the self-consistent potential are also computed within 
\texttt{Quantum-ESPRESSO} for phonons in the irreducible wedge of 
the Brillouin zone. When choosing initial electron and phonon grids, 
it is necessary that the Brillouin zone grid for phonons be 
comensurate with the Brillouin zone grid for electrons in order to map the 
wavefunctions $\psi_{m{\bf k+q}}({\bf r})$ onto $\psi_{m\bf k^\prime+G}({\bf r})$, 
with ${\bf G}$ a reciprocal lattice vector.  For example, if the 
calculation is performed using a Brillouin zone grid of size 
$6\times 6 \times 6$ for the phonons, then the natural choices 
for the electronic Brillouin zone grid are either $6\times 6 \times 6$ or 
$12\times 12 \times 12$. This does not represent a computational 
bottleneck as phonon calculations are considerably more time-consuming
than non-selfconsistent electronic calculations.

\section{Computational methodology}

\subsection{Physical quantities (\texttt{selfen\_phon}, 
\texttt{selfen\_elec}, \texttt{a2f})} \label{sec.physics}

In this section we describe some of the physical quantities 
which can be calculated using \texttt{EPW}. 
The imaginary part of the phonon self-energy within the Migdal 
approximation \cite{migdal.jetp,allenmikovic} is calculated as:
  %Eqn 2 of diamond PRL with Im() prefix and dropped the ''2'' included in w_k 
  \begin{equation}
  \label{phonselfenergy}
  \Pi^{\prime\prime}_{{\bf q}\nu} =   
  {\rm Im} \sum_{mn,{\bf k}} w_{\bf k} |g_{mn}^{\nu}({\bf k,q})|^2 
  \frac{ f(\epsilon_{n{\bf k}}) - f(\epsilon_{m{\bf k+q}}) }{
  \epsilon_{n{\bf k}} - \epsilon_{m{\bf k+q}} - \omega_{{\bf q}\nu} 
  + i\eta}.
  \end{equation}
In Eq.\ (\ref{phonselfenergy}) the electron-phonon matrix elements are given by
  %Eqn 3 of diamond PRL
  \begin{equation}
  \label{eqn_g}
  g_{mn}^{\nu}({\bf k},{\bf q}) =  \langle \psi_{m{\bf k+q}} | 
  \partial_{{\bf q}\nu}V | \psi_{n{\bf k}}\rangle,
  \end{equation}
with $\psi_{n{\bf k}}$ the electronic wavefunction for band $m$, wavevector 
${\bf k}$, and eigenvalue $\epsilon_{n{\bf k}}$, $\partial_{{\bf q}\nu}V$ 
the derivative of the self-consistent potential associated with a phonon 
of wavevector ${\bf q}$, branch index $\nu$, and frequency 
$\omega_{{\bf q}\nu}$. The factors $f(\epsilon_{n{\bf k}}), 
f(\epsilon_{m{\bf k+q}})$ in Eq.\ (\ref{phonselfenergy}) are the Fermi 
occupations, and $w_{\bf k}$ are the weights of the ${\bf k}$-points 
normalized to 2 in order to account for the spin degeneracy in spin-unpolarized 
calculations.  A very common approximation to Eq.\ (\ref{phonselfenergy}) 
consists of neglecting the phonon frequencies $\omega_{{\bf q}\nu}$
and taking the limit of small broadening $\eta$. The final expression is 
positive definite and is often referred to as the ``double-delta function'' 
approximation \cite{allenmikovic}. This approximation is no longer 
necessary when using \texttt{EPW}. We note that the imaginary part 
of the phonon self-energy in Eq.\ (\ref{phonselfenergy}) also corresponds 
to the phonon half-width at half-maximum $\gamma_{{\bf q}\nu}$.

The electron-phonon coupling strength associated with a specific
phonon mode and wavevector $\lambda_{{\bf q}\nu}$ is given by
  \begin{equation}
  \label{lambdaq}
  \lambda_{{\bf q}\nu} = 
   \frac{1}{N_{\rm F}\omega_{{\bf q}\nu}}\sum_{mn,{\bf k}} 
  w_{{\bf k}}|g_{mn}^\nu({\bf k},{\bf q})|^2 \\
  \delta(\epsilon_{n{\bf k}})\delta(\epsilon_{m{\bf k}+{\bf q}}),
  \end{equation}
with $\delta$ being the Dirac delta function. In the double-delta function
approximation the coupling strength $\lambda_{{\bf q}\nu}$ can be related 
to the imaginary part of the phonon self-energy $\Pi^{\prime\prime}_{{\bf q}\nu}$  
as follows:
  \begin{equation}
  \label{lambda_approx}
  \lambda_{{\bf q}\nu} = \frac{1}{\pi N_{\rm F}} 
 \frac{\Pi^{\prime\prime}_{{\bf q}\nu}}{\omega^2_{{\bf q}\nu}}
  \end{equation}
The total electron-phonon coupling $\lambda$ is calculated as 
the Brillouin-zone average of the mode-resolved coupling strengths 
$\lambda_{{\bf q}\nu}$:
  \begin{equation}
  \label{lambda}
  \lambda = \sum_{{\bf q}\nu} w_{{\bf q}} \lambda_{{\bf q}\nu}.
  \end{equation}
In Eq.\ (\ref{lambda}) the $w_{\bf q}$ are the Brillouin zone weights 
associated with the phonon wavevectors ${\bf q}$, normalized to 1 in 
the Brillouin zone. The Eliashberg spectral function $\alpha^2 F$ can 
be calculated in terms of the mode-resolved coupling strengths 
$\lambda_{{\bf q}\nu}$ and the phonon frequencies using:
  % checked that 2 int a2F/w dw = lambda --> self consistent
  \begin{equation}
  \label{a2f}
  \alpha^2F(\omega) = \frac{1}{2}\sum_{{\bf q}\nu}  
  w_{{\bf q}} \omega_{{\bf q}\nu} \lambda_{{\bf q}\nu} \, \delta( \omega - \omega_{{\bf q}\nu}).
  \end{equation}
The transport spectral function $\alpha^2 F_{\rm T}$ is obtained from 
the Eliashberg spectral function $\alpha^2F$ by replacing 
$\lambda_{{\bf q}\nu}$ with $\lambda_{{\rm T},{\bf q}\nu}$:
  \begin{equation}
  \label{a2ftr}
  \alpha^2F_{\rm T}(\omega) = \frac{1}{2}\sum_{{\bf q}\nu}   \\
  w_{{\bf q}} \omega_{{\bf q}\nu} \lambda_{{\rm T},{\bf q}\nu} 
  \delta(\omega - \omega_{{\bf q}\nu}),
  \end{equation}
  % the same as lambda, but with 1- (vk dot vkkq)/|vk|^2 inside the sum
  % same as Grimvall 8.20 but in the form of Eqn.3 above
  \begin{equation}
  \label{lambda_tr}
  \lambda_{{\rm T},{\bf q}\nu} =
  \frac{1}{N_{\rm F}\omega_{{\bf q}\nu}}\sum_{mn,{\bf k}}
  w_{{\bf k}}|g_{mn}^\nu({\bf k},{\bf q})|^2 \\
  \delta(\epsilon_{n{\bf k}})\delta(\epsilon_{m{\bf k}+{\bf q}})
  \left (1 - \frac{{\bf v}_{n{\bf k}} \cdot {\bf v}_{m{\bf k+q}}}{ |{\bf v}_{n{\bf k}}|^2}\right),
  \end{equation}
with ${\bf v}_{n{\bf k}} = \nabla_{\bf k}\epsilon_{n{\bf k}}$ the electron velocity.

The real and imaginary parts of the electron self-energy 
%$\Sigma_{n{\bf k}}^{\prime \prime}$ 
$\Sigma_{n{\bf k}} = \Sigma_{n{\bf k}}^{\prime} + i\Sigma_{n{\bf k}}^{\prime \prime} $ 
can be calculated as
  % Eqn 5.42 of Grimvall
  \begin{equation}
  \label{elselfenergy}
  \Sigma^{}_{n{\bf k}} = 2 
  \sum_{{\bf q}\nu} w_{{\bf q}} 
  |g_{mn}^\nu({\bf k,q})|^2 
  \left[ \frac{n(\omega_{{\bf q}\nu})+ 
  f(\epsilon_{m{\bf k+q}})}{\epsilon_{n{\bf k}} - \epsilon_{m{\bf k+q}} 
  + \omega_{{\bf q}\nu} - i\eta}  + 
  \frac{n(\omega_{{\bf q}\nu})+ 1 
  -f(\epsilon_{m{\bf k+q}})}{\epsilon_{n{\bf k}} - \epsilon_{m{\bf k+q}} 
  - \omega_{{\bf q}\nu} +i\eta}  \right],
  \end{equation}
with $n(\omega_{{\bf q}\nu})$ the Bose occupation factors.

\subsection{Calculation of the matrix elements on the coarse Brillouin zone grid
(\texttt{elphon\_shuffle\_wrap})}

The key task of \texttt{EPW} is to calculate from first-principles
electron-phonon matrix elements for a large number of electron and
phonon wavevectors with a modest computational effort. The initial step 
of this process is to determine the electron-phonon matrix elements 
on coarse grids of electron and phonon wavevectors using standard DFT 
and DFPT methods \cite{el-ph_bloch}.
The result of this step is a set of matrix elements given by 
Eq.\ (\ref{eqn_g}). The wavefunctions and the phonon perturbation 
potentials are read into \texttt{EPW}, then the matrix elements on the 
coarse grid are calculated within \texttt{EPW}.

The phonon dynamical matrices and the
linear variations of the self-consistent potential are read
from file only for wavevectors in the irreducible wedge of the
Brillouin zone. The dynamical matrices and the potentials variations
for all the remaining points of the uniform grid are generated
using crystal symmetry operations.  
This strategy is advantageous since the computation of the phonon
dynamical matrices and potential variations is generally the most 
time-consuming part of an electron-phonon calculation. 
For example, a system with cubic symmetry requiring a coarse mesh 
of $8\times 8 \times 8$ phonon wavevectors needs only have phonons 
calculated at 29 irreducible points; hence the reduction in computational 
time is a factor of 512/29.

The electron-phonon matrix elements and the dynamical matrices 
thus calculated at each point of the coarse Brillouin-zone
grid can be written to disk through an input file option
(\texttt{prefix.epb} files). Subsequent executions of 
\texttt{EPW} can forgo the recalculation of the  matrix elements 
and dynamical matrices by reading these data from disk. 

\subsection{Interface with \texttt{wannier90} (\texttt{pw2wan90epw})}

\texttt{wannier90} generates maximally localized Wannier functions 
by minimizing the spread of the Berry-phase position operator
through a unitary transform \cite{marzari.wannier}. Details on 
how to run \texttt{wannier90} can be found in the \texttt{wannier90} 
documentation \cite{w90}.  

In standalone mode \texttt{wannier90} reads three files generated
from a first-principles calculation (\texttt{prefix.mmn, prefix.amn, 
prefix.eig}), a file which defines the starting guess for determining
MLWFs and the crystal structure (\texttt{prefix.nnkp}), and a runtime 
input file (\texttt{prefix.win}). The execution of \texttt{wannier90} 
therefore requires the user to run multiple programs and handle the 
files to be passed between \texttt{wannier90} and \texttt{Quantum-ESPRESSO}. 
In order to simplify the calculation of the electron-phonon coupling
\texttt{EPW} calls \texttt{wannier90} as a library. \texttt{EPW} therefore
requires only the wavefunction files from \texttt{Quantum-ESPRESSO}
and a runtime input file in order to determine MLWFs using \texttt{wannier90}.
The quantites required for running \texttt{wannier90} are either 
calculated within \texttt{EPW} (for instance the overlap matrices 
$A_{mn}$ and $M_{mn}$ \cite{w90}), or else read in from file (for 
instance the Kohn-Sham eigenvalues). These data are then passed from 
\texttt{EPW} to \texttt{wannier90} through the \texttt{wannier\_run} 
library routine. This feature of \texttt{EPW} ensures that the execution 
of \texttt{wannier90} is embedded within \texttt{EPW}. Hence the end user 
is only required to run one single executable which communicates directly 
with \texttt{wannier90}.  \texttt{EPW} also includes an option to pass 
additional input parameters to \texttt{wannier90}. This allows the user 
to access all the available features of \texttt{wannier90}, such as for 
instance plotting bandstructures or MLWFs.  When called from \texttt{EPW}, 
\texttt{wannier90} produces the gauge matrix $U^{\bf k}_{mn}$ \cite{w90}, 
which yields the transformation between Bloch eigenstates and MWLFs, according to:
  %% Eqn 1 of wannier90 CPC
  \begin{equation}
  \label{eqn_u}
  w_{n{\bf R}_e}({\bf r}) = \frac{\Omega\,\,}{(2\pi)^3}\int_{\rm BZ} d{\bf k} \ {\rm e}^{-i {\bf k} \cdot {\bf R}_e }
  \sum_m U^{{\bf k}}_{mn} \, \psi_{m{\bf k}}({\bf r}),  
  \end{equation}
where $w_{n{\bf R}_e}$ is a MLWF associated with the direct lattice 
vector ${\bf R}_e$, $\Omega$ is the unit cell volume, and the integral 
is discretized over the Brillouin zone.  The array $U^{\bf k}_{mn}$ 
has the dimensions of the number of Bloch bands times the number of 
MLWFs times the number of electronic wavevectors in the coarse Brillouin 
zone grid. This matrix is written to disk and can directly be read on 
subsequent program executions. The array $U^{\bf k}_{mn}$ is used in 
\texttt{EPW} in order to transform Bloch functions into MLWFs.

\subsection{Transformation from the Bloch representation on the coarse 
Brillouin zone grid to the MLWF representation (\texttt{ephwann\_shuffle})}

After calculating the electron-phonon matrix elements in the Bloch 
representation for each wavevector on the coarse electron and phonon 
Brillouin zone grids, \texttt{EPW} transforms the electronic Hamiltonian, 
the phonon dynamical matrix, and the electron-phonon matrix elements into
the Wannier representation. Derivations and detailed explanations of the 
following transformations can be found in Ref.~\cite{giustino.method}.
For clarity the electron band and the phonon mode indices will be omitted 
in the following equations. The electronic Hamiltonian in the MLWF 
representation $H^{\rm el}_{{\bf R}_{\rm e},{\bf R}^\prime_{\rm e}}$  is obtained as:
  %% Eqn 26 of EPW PRB
  \begin{equation}
  \label{ham}
  H^{\rm el}_{{\bf R}_{\rm e},{\bf R}^\prime_{\rm e}} =
   \sum_{\bf k} w_{\bf k} {\rm e}^{-i {\bf k} \cdot ({\bf R}^\prime_{\rm e}-{\bf R}_{\rm e}) }
  U^\dagger_{\bf k} H^{\rm el}_{\bf k} U_{\bf k}.
  \end{equation}
In this case the weights $w_{\bf k}$ are normalized to 1.
The Hamiltonian matrix elements in the Wannier representation 
$H^{\rm el}_{{\bf R}_{\rm e},{\bf R}^\prime_{\rm e}}$ decay rapidly with the distance 
$|{\bf R}_{\rm e}-{\bf R}^\prime_{\rm e}|$, as they scale with the overlap of MLWFs 
centered in the unit cells ${\bf R}={\bf R}_{\rm e}$ and ${\bf R}={\bf R}^\prime_{\rm e}$, 
respectively. The dynamical matrix can be transformed to a localized 
real-space representation using 
  %% Eqn 29 of EPW PRB
  \begin{equation}
  \label{dyn}
  D^{\rm ph}_{{\bf R}_{\rm p},{\bf R}^\prime_{\rm p}}=
  \sum_{\bf q} w_{\bf q}
  {\rm e}^{-i {\bf q} \cdot ({\bf R}^\prime_{\rm p}-{\bf R}_{\rm p})}
  {\bf e}_{\bf q} D^{\rm ph}_{\bf q} {\bf e}^\dagger_{\bf q},
  \end{equation}
where ${\bf e}_{\bf q}$ are the orthonormal eigenvectors of the dynamical 
matrix. The matrix $D^{ph}_{{\bf R}_{\rm p},{\bf R}^\prime_{\rm p}}$ is the mass-scaled 
matrix of force constants. The electron-phonon matrix elements in 
the MLWF representation are obtained using:
  %% Eqn 24 of EPW PRB
  \begin{equation}
  \label{g}
  g({\bf R}_{\rm e}, {\bf R}_{\rm p}) = 
  \frac{1}{N_{\rm p}} \sum_{\bf k,q} w_{\bf k} w_{\bf q}
  {\rm e}^{-i\left({\bf k}\cdot {\bf R}_{\rm e} + {\bf q}\cdot {\bf R}_{\rm p}\right)} 
  U^\dagger_{\bf k+q} g({\bf k,q})U_{\bf k} {\bf u}_{\bf q}^{-1}
  \end{equation}
where the ${\bf u}^\nu_{{\bf q}\kappa}$ are the phonon eigenvectors 
scaled by the atomic masses \cite{giustino.method}. In order to check 
the spatial decay of $H^{\rm el}_{{\bf R}_{\rm e},0}$,
$D^{\rm ph}_{{\bf R}_{\rm p},0}$, and $g({\bf R}_{\rm e}, {\bf R}_{\rm p})$
the magnitude of these quantites as a function of ${\bf R}_{\rm e}$ and ${\bf R}_{\rm p}$ 
is written to formatted files in the working directory. A run-time 
option allows for all data in the Wannier representation to be written 
into one single file. For subsequent calculations these data can be 
read in and program execution can restart without the need to go through 
the prior computational steps.

\subsection{Tranformation from the MLWF representation to the Bloch representation 
on the fine Brilluin zone grid (\texttt{ephwann\_shuffle})}
\label{sec.fourier}

The accuracy of \texttt{EPW} calculations depends on the spatial 
localization of the MLWFs and the phonon perturbations.
Typically MLWFs are localized within a few \AA \cite{localization, localization2}.  
The localization of the phonon perturbation is dependent upon 
the dielectric screening properties of the system, and must be verified 
in each case before proceeding with the interpolation \cite{giustino.method}. 

In order to calculate the electronic eigenstates, the phonon frequencies,
and the electron-phonon matrix elements on a fine Brillouin zone grid,
the Hamiltonian, the dynamical matrix, and the electron-phonon matrix 
elements are truncated outside of a 
real-space supercell containing $N_{\bf k}$ and $N_{\bf q}$
unit cells in the case of electrons and phonons, respectively. 
Here $N_{\bf k}$ and $N_{\bf q}$
are the number of grid points in the coarse Brillouin zone meshes.
Following this truncation it is possible to perform an interpolation back into the
Bloch representation onto arbitrary electron and phonon wavevectors.
The Hamiltonian, the dynamical matrix, 
and the electron-phonon matrix elements are
Fourier-transformed back to the Bloch representation
by inverting Eqs. (\ref{ham}), (\ref{dyn}), and (\ref{g}) \cite{giustino.method}.

The procedure described here enables the calculation of electron-phonon matrix 
elements on extremely fine Brillouin zone grids. The fine grids of electron
and phonon wavevectors are specified by the user in input.
The procedure adopted here is similar to the Fourier interpolation 
of time series commonly adopted in signal processing \cite{Fourier}.
The same strategy as the one employed in \texttt{EPW} 
has been adopted in order to study Fermi surfaces
\cite{PhysRevB.75.195121} and the anomalous Hall effect \cite{PhysRevB.74.195118}.

At this stage the Hamiltonian, the dynamical matrix, and the electron-phonon 
matrix elements on fine grids of electron and phonon wavevectors are used 
to compute the physical quantities described in Sec.\ \ref{sec.physics}.

\subsection{Summary of the tasks executed by \texttt{EPW}}
The computational steps described in the previous sections can be summarized 
schematically as follows:
\begin{description}
\item[1.]{Electron eigenstates and eigenvalues are read from disk,} 
\item[2.]{\texttt{wannier90} input data including the overlap 
 matrices $A_{mn}$ and $M_{mn}$ are computed,}
\item[3.]{\texttt{wannier90} is called as a library, and the resulting 
  MLWF localization matrix is stored on disk.}
\item[4.]{The electron-phonon matrix elements are computed on a coarse grids 
 of electron and phonon wavevectors in the Bloch representation,}
\item[5.]{The Hamiltonian, dynamical matrix, and electron-phonon matrix elements
 are transformed from the Bloch representation to the MLWF representation,}
\item[6.]{The Hamiltonian, dynamical matrix, and electron-phonon matrix elements
 are Fourier-transformed back to the Bloch representation 
  on arbitrarily dense grids of electron and phonon wavevectors,}
\item[7.]{The electronic eigenvalues, phonon frequencies, and 
  electron-phonon matrix elements are processed in order to calculate
  physical quantites such as, for instance,
  the electron self-energy or the electron-phonon coupling strength.}
\end{description}

\section{Parallelization}

\texttt{EPW} is designed and optimized to be executed on multiple processors.  
Within the \texttt{Quantum-ESPRESSO} package, parallel tasks are split into 
processor ``pools'', where each pool is allocated a set of electronic 
wavevectors {\bf k}.  If there is more than one processor within each pool, 
the reciprocal space {\bf G}-vectors for describing the wavefunctions are 
split amongst the processors of that pool. \texttt{EPW} can be executed in 
parallel by splitting the calculation over electron or phonon wavevectors. 
\texttt{EPW} is not parallelized over the reciprocal-space {\bf G}-vectors.
The parallelization strategy in \texttt{EPW} is tailored to each step of 
the computation.  As execution of \texttt{EPW} begins, the coarse 
{\bf k}-point mesh is generated and distributed amongst the available 
pools, as in \texttt{Quantum-ESPRESSO}. 

Most of the inputs to the \texttt{wannier90} such as the lattice and 
reciprocal vectors are passed through \texttt{Quantum-ESPRESSO} and 
\texttt{EPW}. Important exceptions are the overlap matrices $A_{mn}$ 
and $M_{mn}$.  Within \texttt{EPW} the computation of the overlap 
matrix elements is distributed across the coarse grid {\bf k}-point pools.
As the elements of the matrices $A_{mn}$ and $M_{mn}$ are computed 
independently, the execution time of this step scales inversely with 
the the number of pools.

\texttt{wannier90} is executed serially on each processor through a 
library call. The time required to determine MLWFs is negligible as 
compared to the computation of the electron-phonon matrix elements 
on the coarse Brillouin zone grids.

The electron-phonon matrix elements on the coarse Brillouin zone grid 
are computed sequentially for each phonon wavevector, while the 
electron wavevectors are distributed across pools.

The computation of the electron-phonon matrix elements on the fine 
Brillouin zone grids is parallelized over electronic or phonon 
wavevectors depending on the calculation type. For example, when 
calculating the phonon self-energy an integration is required over 
electronic wavevectors. In this case the most convenient strategy
is to parallelize over the electronic wavevectors. The converse is 
true for the calculation of the electron self-energy. In those cases 
where the final quantity requires integrations over both electron 
and phonon wavevectors, such as the total electron-phonon coupling 
strength, the efficiency of \texttt{EPW} is independent of the choice 
of the parallelization scheme.

\begin{figure}
 \includegraphics[width=0.99\columnwidth]{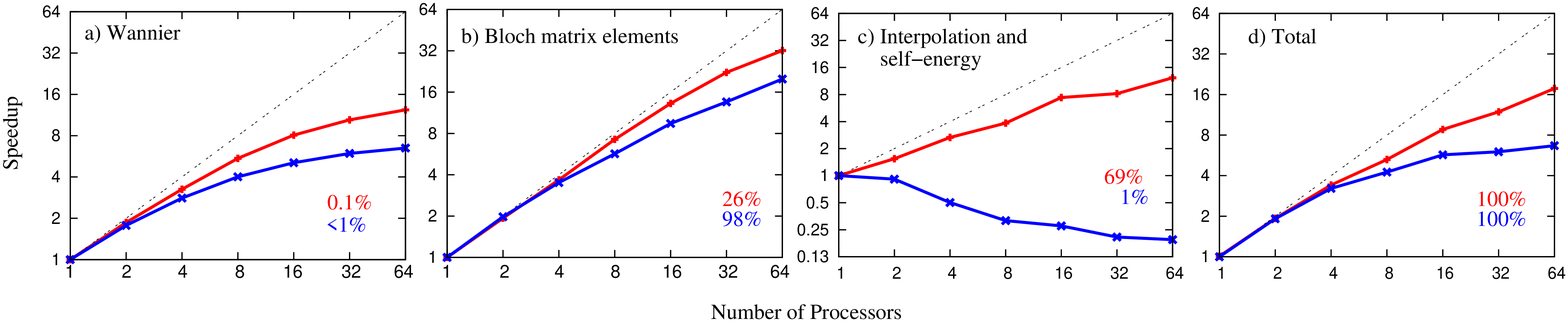}
  \caption{
  \label{fig1_log}
  Parallelization of \texttt{EPW}: speedup vs number of processors observed 
  in parallelizing a) the generation of MLWFs, b) the calculation of the 
  electron-phonon matrix elements on the coarse Brillouin zone grids,
  c) the interpolation of the electron-phonon matrix elements to the fine
  Brillouin zone grids as well as the computation of the phonon self-energy. 
  Panel d) displays the speedup observed to perform the entire computation 
  including the initialization step and the intermediate I/O.
  The speedup is defined as the ratio between the time it takes to run 
  a calculation on a single processor and the time it takes to run the same
  calculation on a given number of processors.  
  The diagonal dotted lines correspond to the ideal speedup, which is equal
  to  the number of processors employed.  
  The calculations have been performed for hole-doped SiC in the rigid-band
  approximation. The unit cell contains 2 atoms.
  The blue line corresponds to a calculation with coarse Brillouin zone
  grids containing $6\times 6\times 6$ ${\bf k}$- and ${\bf q}$-points, 
  a kinetic energy cutoff of 30 Ry, and fine Brillouin zone grids containing 
  1000 points each.
  The red line corresponds to a calculation with coarse Brillouin zone
  grids containing $8\times 8\times 8$ ${\bf k}$- and ${\bf q}$-points, 
  a kinetic energy cutoff of 60 Ry, and fine ${\bf k}$- and ${\bf q}$-points grids 
  containing $10^6$ and 104 points (the irreducible
  points of a uniform and unshifted $14\times 14\times 14$ grid), respectively. 
  The parallelization becomes more efficient
  as the number of electron and phonon wavevectors in the fine
  Brillouin zone grids is increased.
  }
  \end{figure}

Figure \ref{fig1_log} shows the efficiency of our parallelization 
strategy for various sections of \texttt{EPW} for the test case of 
hole-doped SiC. This example is included for reference in the \texttt{EPW} 
distribution.

\subsection{Calculations ``on the fly''}

The interpolation part of \texttt{EPW} can be executed in two different modes.
In the first mode \texttt{EPW} calculates all the electron-phonon matrix
elements for every {\bf k}- and {\bf q}-point of the fine Brillouin zone grids, 
and stores this information to disk. The subroutines for calculating the
electron or phonon self-energy are then called and the matrix elements along 
with the electronic eigenvalues and the phonon frequencies are read from disk
in order to evaluate Eqs.\ (\ref{phonselfenergy}), (\ref{elselfenergy}).

In the second mode of operation, the electron or the phonon self-energy are
calculated ``on the fly'' as the matrix elements are generated, and all the
quantities are overwritten. In this case the matrix elements are not stored on disk.

This feature allows us to address systems for which the storage required
for the electron-phonon matrix elements would be exceedingly large. For example,
if the fine electron and phonon Brillouin zone grids consisted each of 
$10^5$ wavevectors, and we had 8 MLWFs and 6 phonon modes, then
the entire set of double precision matrix elements would require more than 1 TB
of storage. On the other hand the calculation on the fly would reduce 
the storage requirement down to a manageable size of 1 TB.

\section{Examples}

In the following we illustrate the capabilities of \texttt{EPW} by 
describing three prototypical systems: (i) Lead, a simple metal 
which is also a strong-coupling superconductor. (ii) Graphene, a 
two-dimensional zero-gap semiconductor with linear quasiparticle 
dispersions close to the Fermi level. (iii) Silicon carbide, 
a wide-gap semiconductor which becomes a low-temperature 
superconductor upon boron doping.

\subsection{Lead}

Solid Pb is the prototypical strong-coupling superconductor. While 
the superconducting transition temperature of Pb is less than 10 K, 
the electron-phonon coupling strength has been measured and calculated 
to be very large, in the range of $\lambda = 1.3\pm 0.1$ 
\cite{mcmillan.rowell,pb.refs.cohen,pb.refs.barth,pb.refs.franck,noffsinger.pblayer}.

In this section we present an example calculation of the electron-phonon 
coupling in bulk Pb and we compare our results to experimental data.
The calculations were performed within the local-density approximation 
to density functional theory using \texttt{Quantum-ESPRESSO}, 
\texttt{wannier90}, and \texttt{EPW}.  A norm-conserving, 
scalar-relativistic pseudopotential including four valence electrons 
was used to take into account the core-valence interaction. The 
electronic states were computed within a plane-wave basis with a kinetic 
energy cutoff of 80 Ry.  The charge density was computed self-consisently 
on a $16 \times 16 \times 16$ $\Gamma$-centered Brillouin zone grid. The 
electronic wavefunctions used within \texttt{EPW} were calculated 
on an $8 \times 8 \times 8$ uniform Brillouin zone grid. The phonon 
dynamical matrices and the linear variations of the self-consistent 
potential were calculated within DFPT using the \texttt{Quantum-ESPRESSO} 
package, using the same convergence parameters as above. We considered 
a uniform $8 \times 8 \times 8$ Brillouin zone grid for the phonon 
calculations, corresponding to 29 irreducible phonon wavevectors.

\begin{figure}
  \includegraphics[width=0.7\columnwidth]{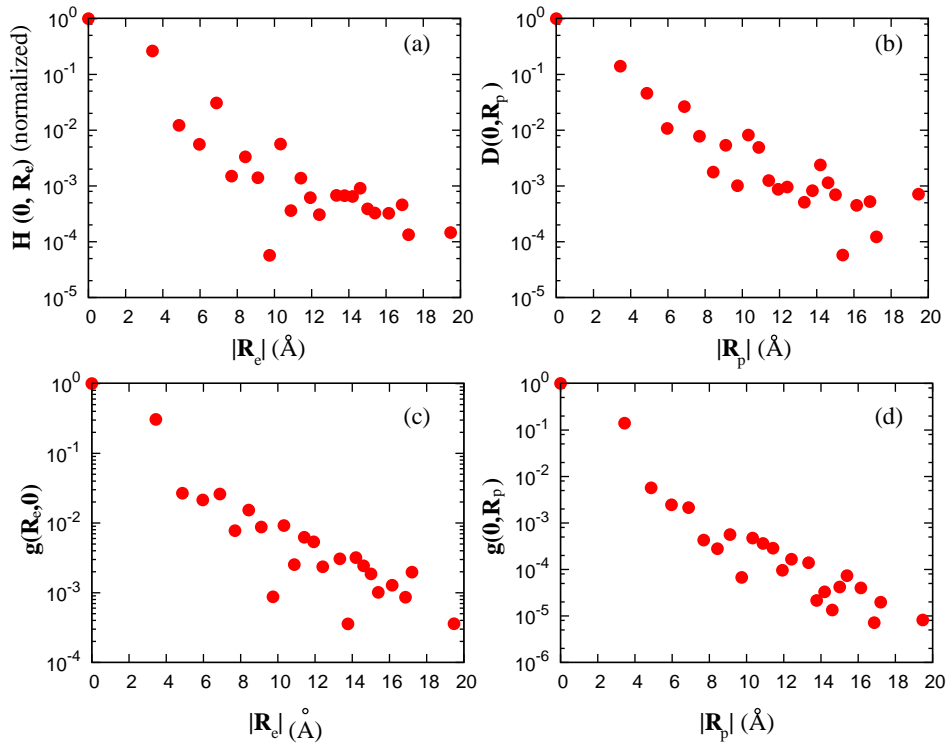}
  \caption{
  \label{fig.pb.decay}
  Spatial decay of the largest components of the Hamiltonian 
  $H^{\rm el}_{{\bf R}_{\rm e},0}$ (a), the dynamical matrix 
  $D^{\rm ph}_{{\bf R}_{\rm p},0}$ (b), and the electron-phonon matrix
  elements $g(0,{\bf R}_{\rm p})$ (c) and $g({\bf R}_{\rm e},0)$ 
  (d) for fcc Pb. The data are plotted as a function of distance 
  and are taken along several directions.
  }
  \end{figure}

Four Wannier states were used to reconstruct the electronic structure 
near the Fermi level. These states were found to be \textit{$sp^3$}-like 
functions localized symmetrically along each of the Pb-Pb bonds, with a 
spatial spread of 2.17 \AA.  The spatial decay of the electron Hamiltonian, 
the phonon dynamical matrix, and the electron-phonon matrix elements in 
the MLWF representation are shown in Fig. \ref{fig.pb.decay}.  

\begin{figure}
  \includegraphics[width=0.5\columnwidth]{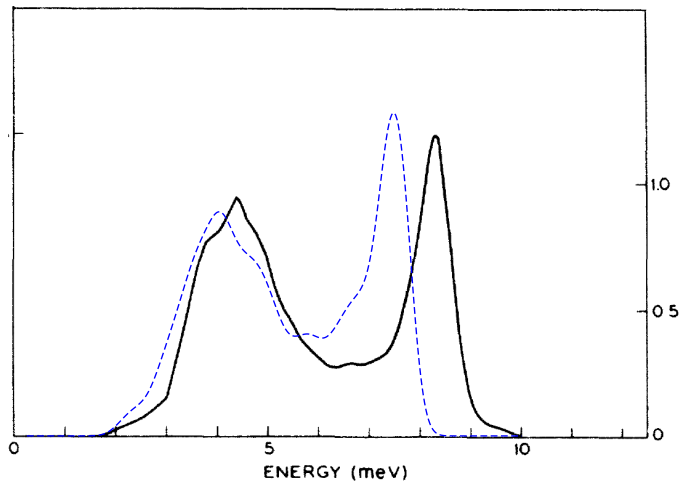}
  \caption{
  \label{a2fpb}
  Eliashberg spectral function of fcc Pb.  The solid line is from Ref. \cite{mcmillan.rowell},
  the dashed line has been obtained using \texttt{EPW}.
  }
  \end{figure}

In order to calculate the phonon linewidths and the total electron-phonon 
coupling, the electron Hamiltonian, the phonon dynamical matrix, and the 
electron-phonon matrix elements were transformed from the MLWF 
representation to the Bloch representation by the generalized Fourier 
interpolation described in Sec.\ \ref{sec.fourier}. By using $10^6$ 
{\bf k}-points, 8000 {\bf q}-points, and a smearing parameter of 30 meV, 
we obtained a total electron-phonon coupling strength $\lambda =$~1.41.
The Eliashberg spectral function $\alpha^2 F(\omega)$ is shown in 
Fig.\ \ref{a2fpb}, together with the experimental curve obtained by 
inverting tunneling data \cite{mcmillan.rowell}. Our method clearly 
yields a very good agreement with experiment.

For comparison with experiment we also calculated the superconducting 
transition temperature $T_{\rm c}$ starting from the total electron-phonon 
coupling.  We adopted the modified McMillan equation \cite{allen.dynes} 
and a Coulomb repulsion parameter $\mu^*$ between 0.10 and 0.15, 
obtaining $T_c$ equal to 4K to 6K respectively. This result compares favorably to
the experimental value 7.2 K.

\subsection{Graphene}

Graphene is a two-dimensional sheet of carbon atoms which has attracted
enourmous attention from the scientific community due to its peculiar
electronic properties \cite{graphene.fab}, and in particular
the Dirac-like nature of electrons near the Fermi 
level \cite{dirac.nature,dirac.nature2}. Considerable attention 
has been paid to the measurement of many-body renormalization effects, 
such as the effect of the electron-phonon interaction on the electronic 
bandstructure, using for instance angle-resolved photoemission
experiments \cite{rotenberg}.

We here present the results of an electron-phonon calculation on a sheet of 
freestanding graphene \cite{PhysRevLett.99.086804}. 
The local density approximation to density 
functional theory was employed in conjunction with a norm-conserving 
carbon pseudopotential. A plane wave basis with a kinetic energy 
cutoff of 60 Ry was used.  The graphene sheets were separated by 
a vacuum of 8 \AA\ in a supercell geometry in order to eliminate 
spurious interactions between periodic replicas. The relaxed C-C 
bond length was found to be 1.405 \AA.

\begin{figure}
  \includegraphics[width=0.7\columnwidth]{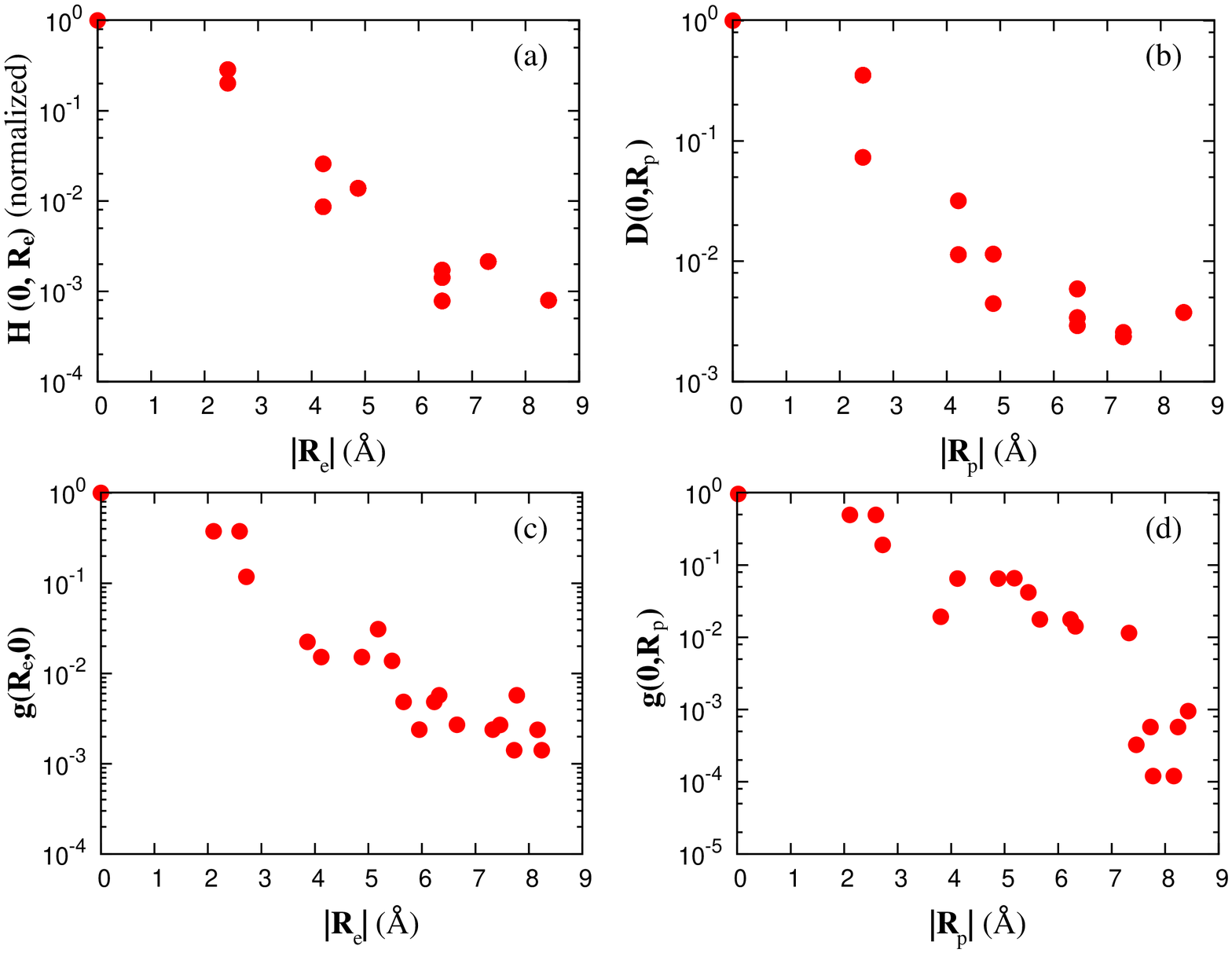}
  \caption{
  \label{fig.graphene.decay}
  Spatial decay of the largest components of the Hamiltonian 
  $H^{\rm el}_{{\bf R}_{\rm e},0}$ (a), the dynamical matrix 
  $D^{\rm ph}_{{\bf R}_{\rm p},0}$ (b), and the electron-phonon matrix
  elements $g(0,{\bf R}_{\rm p})$ (c) and $g({\bf R}_{\rm e},0)$ 
  (d) for graphene. The data are plotted as a function of distance 
  and are taken along several directions.
  }
  \end{figure}

Three in-plane bonding Wannier functions, two $p_z$-like Wannier functions (one per each
carbon atom), and two $s$-like Wannier functions directly above and below the center
of the hexagon (away from the graphene plane by 1.57 \AA) in the
two-atom unit cell were used to describe the 
electronic structure. 
The spatial spread of these MLWFs are 0.565 \AA, 0.782 \AA, and
1.726 \AA, respectively.
The spatial decay of the electron Hamiltonian,
the phonon dynamical matrix, and the electron-phonon matrix elements in
the MLWF representation for a $6 \times 6 \times$ 1 grid are shown in Fig. \ref{fig.graphene.decay}.

\begin{figure}
  \includegraphics[width=0.99\columnwidth]{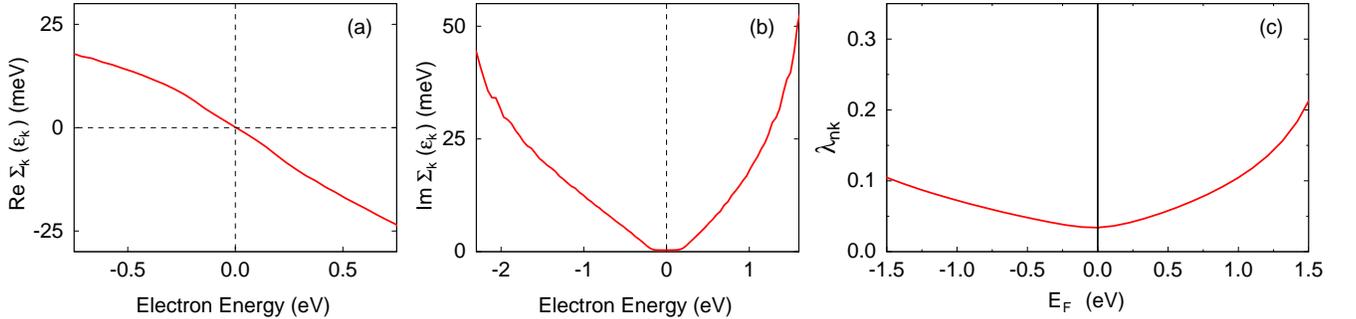}
  \caption{
  \label{graphene.results}
  The real (a) and imaginary (b) parts of the electron self-energy $\Sigma_{\bf k}$ near the Dirac 
  point as well as the electron-phonon coupling $\lambda_{n{\bf k}}$ as a 
  function of doping (c) for graphene. In panels (a) and (b), the Fermi energy has been set to zero.
  }
  \end{figure}

The electron self-energy was computed and the integral of of Eq. 
(\ref{elselfenergy}) was performed with $10^6$ phonon {\bf q}-vectors, 
obtained through the interpolation method of this communication.  
The mass renormalization parameter, $\lambda_{n{\bf k}}$ has been obtained 
as an energy derivative of the electron self-energy as in
$\lambda_{n{\bf k}} = \frac{\partial}{\partial \epsilon}
 {\rm Re}\  \Sigma_{n{\bf k}}(\epsilon_{n{\bf k}}) |_{\epsilon=\epsilon_F}.$
The results of the calculations of the real and imaginary parts of the
electron self-energy as well as the electron-phonon coupling parameter as a
function of Fermi energy are given in Fig. \ref{graphene.results}.

\subsection{Silicon carbide}

The possibility of achieving superconducting behavior in semiconductors has 
recently attracted considerable interest, and several experimental 
\cite{semi.expt, diamond.expt} and theoretical \cite{blase.sic, 
noffsinger.sic, cohen.rmp, giustino.diamond,PhysRevLett.93.237003,
PhysRevLett.93.237002,PhysRevB.72.014306,PhysRevLett.93.237004} studies 
have been performed on this class of materials. Boron-doped SiC is a 
promising candidate because of its potential uses in power electronics 
owing to its large breakdown voltage.

\begin{figure}
  \includegraphics[width=0.7\columnwidth]{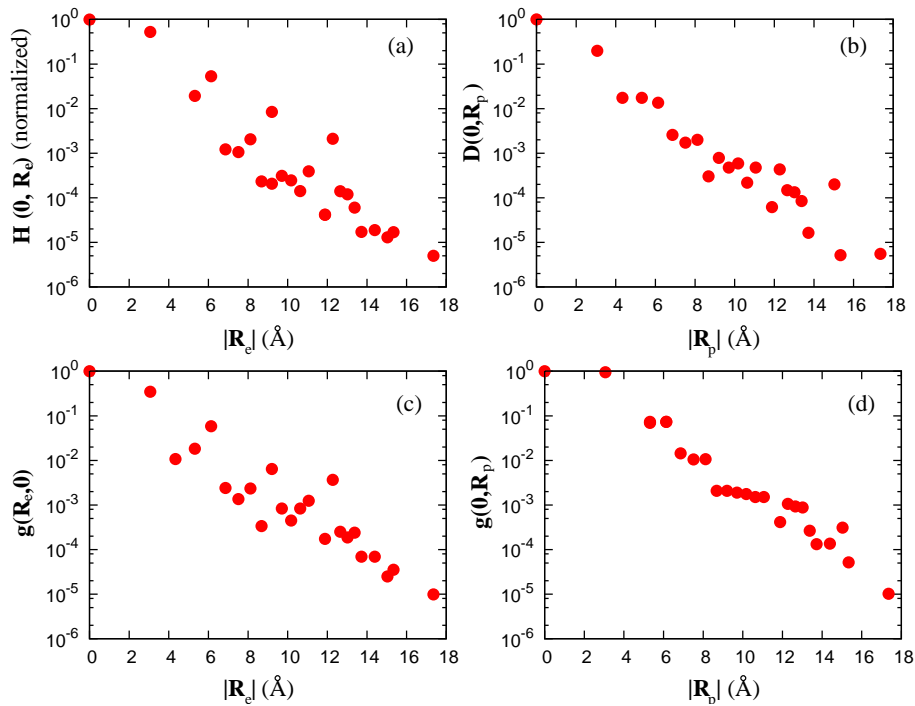}
  \caption{
  \label{figdecay}
  Spatial decay of the largest components of the Hamiltonian 
  $H^{\rm el}_{{\bf R}_{\rm e},0}$ (a), the dynamical matrix 
  $D^{\rm ph}_{{\bf R}_{\rm p},0}$ (b), and the electron-phonon matrix
  elements $g(0,{\bf R}_{\rm p})$ (c) and $g({\bf R}_{\rm e},0)$ (d) 
  for cubic hole-doped SiC. The data are plotted as a function of 
  distance and are taken along several directions.}
  \end{figure}

For this test case we performed electron-phonon calculations on a rigid-band
model of 4\% hole-doped cubic SiC. For our calculations we employed the local 
density approximation to density functional theory, norm-conserving 
pseudopotentials, and a plane wave basis with a kinetic energy cutoff of 
60 Ry. The relaxed lattice parameter was found to be 4.34 \AA, in good 
agreement with the experimental value of 4.36 \AA. \texttt{EPW} was executed 
using coarse meshes of $8\times$ 8 $\times 8$ uniform grids of electron and 
phonon wavevectors. Four Wannier functions per formula unit were considered
to describe the valence electronic struture. MLWFs were found to be
$sp_3$-like functions with an average spread of 1.15 \AA. The spatial 
decay of the electron Hamiltonian, the phonon dynamical matrix, and the 
electron-phonon matrix elements in the MLWF representation are shown in 
Fig. \ref{figdecay}.
We calculated the total electron-phonon coupling strength $\lambda = 0.34$ 
using 250,000 {\bf k}-points and 8000 {\bf q}-points. The superconducting 
transition temperature was calculated using a Coulomb repulsion parameter 
$\mu^\star = 0.1$ and was found to be $T_{\rm c} =$ 1 K. Our result is in 
good agreement with recent experimental data yielding $T_{\rm c} =$ 1.4 K\cite{semi.expt}.

\section{Conclusion}

In this communication we introduced \texttt{EPW}, a computer code for 
calculating the electron-phonon coupling from first-principles using 
density functional perturbation theory and maximally localized Wannier 
functions. \texttt{EPW} enables extremely accurate and highly efficient 
calculations of the mode-resolved and total electron-phonon coupling 
strength, as well as phonon and electron linewidths. The code is distributed 
through the website \texttt{http://epw.org.uk} and is available under the 
terms of the GNU General Public License. Plans are in place to extend 
\texttt{EPW} in order to implement the anisotropic Eliashberg 
theory \cite{nature.mgb2}, the density functional theory for 
superconductors \cite{gross.prl.1,gross.prl.2}, and phonon-assisted optical 
responses \cite{auger}.

\section{Appendix}

\subsection{Specifying a unique gauge for the electronic wavefunctions (\texttt{setphases})}

Since nondegenerate electronic wavefunctions are uniquely defined up 
to a phase, and a set of degenerate wavefunctions can be mixed via a 
unitary matrix, electron-phonon matrix elements are machine dependent.
In some cases it may be convenient to use electron-phonon matrix 
elements outside of \texttt{EPW}, for instance in the study of phonon 
sidebands in excitonic spectra \cite{exciton}, or in the phonon-assisted 
Auger recombination \cite{Lochmann1980553}. In order to meaningfully use 
the electron-phonon matrix elements generated by \texttt{EPW} outside of 
the code it is necessary to define uniquely the phase of each wavefunction 
as well as the way in which degenerate wavefunctions are mixed.

In order to set the gauge of each wavefunction in \texttt{EPW} we proceed as 
follows.  First, we determine the subset of degenerate wavefunctions at each 
wavevector on the coarse mesh.  Then we artificially lift the degeneracies 
by diagonalizing the subset of degenerate states with respect to a small 
external perturbation. In principle the perturbation could take on any form, 
however for convenience \texttt{EPW} employs a nonlinear combination of the 
data found in the \texttt{prefix.dvscf} files. Finally, the Fourier components 
of each wavefunction are scaled by a complex factor $\exp(i\theta)$ in such 
a way that the largest component of each wavefunction is real-valued.  
In this way, the machine dependence of the matrix elements is eliminated. 
This procedure is explained in detail in Ref.\ \cite{giustino.method}.

\section{Acknowledgments}
We are grateful to Arash A. Mostofi, Jonathan R. Yates, 
Ivo Souza, David Vanderbilt, and Nicola Marzari for useful interactions about 
\texttt{wannier90}, Stefano de Gironcoli and Paolo Giannozzi
for interactions on \texttt{Quantum-ESPRESSO}, and John M. Rowell for permission 
to use the figure contained in Fig. \ref{a2fpb}. J.N. and development of \texttt{EPW}
was supported by 
National Science Foundation Grant No. DMR07-05941.  B.D.M. and codes testing was
supported by the Director, Office 
of Science, Office of Basic Energy Sciences, Materials Sciences and Engineering 
Division, U.S. Department of Energy under Contract No. DE-AC02-05CH11231.
C.-H.P. was supported by Office of Naval
Research MURI Grant No. N00014-09-1066.
The research leading to these results has received funding from the 
European Research Council under the European Community's Seventh Framework 
Programme (FP7/2007-2013) / ERC grant agreement no. 239578.

\end{document}